\documentclass[journal]{IEEEtran}
\usepackage{amsmath}
\usepackage{amssymb}
\usepackage{amsthm}
\usepackage{cite}
\usepackage{color}
\usepackage{xcolor}
\usepackage{caption}
\usepackage{epsfig,latexsym}
\usepackage{float}
\usepackage{fancyhdr}
\usepackage{graphicx}
\usepackage{indentfirst}
\usepackage{mdwmath}
\usepackage{mdwtab}
\usepackage{subfigure}
\usepackage{setspace}
\usepackage{times}
\usepackage{url}
\usepackage{multirow}
\usepackage{algorithm}
\usepackage{algorithmic}
\usepackage{verbatim}
\usepackage{booktabs}  

\newtheorem{remark}{Remark}
\newtheorem{theorem}{Theorem}

\newtheorem{lemma}{Lemma}

\newtheorem{corollary}{Corollary}

\begin{document}

\title{\huge{Multi-/Uni-Cast Non-Orthogonal Multiple Access-Based INAC}}

\author{Yaoyu Zhang, Xin Sun, Tianwei Hou,~\IEEEmembership{Member,~IEEE}, Anna Li,~\IEEEmembership{Member,~IEEE}, Sofie Pollin,~\IEEEmembership{Senior Member,~IEEE},
Yuanwei Liu,~\IEEEmembership{Fellow,~IEEE}, and Arumugam Nallanathan,~\IEEEmembership{Fellow,~IEEE}
 
\thanks{This work was supported in part by the Fundamental Research Funds for the Postgraduate Innovation Project under Grant W25YJS00060, in part by the Beijing Natural Science Foundation L232041, and in part by EPSRC grant numbers to acknowledge are EP/W004100/1, EP/W034786/1 and EP/Y037243/1.}
\thanks{Yaoyu Zhang and Xin Sun are with the School of Electronic and Information Engineering, Beijing Jiaotong University, Beijing 100044, China (e-mail: 24110070@bjtu.edu.cn, xsun@bjtu.edu.cn). }
\thanks{Tianwei Hou is with the School of Electronic and Information Engineering, Beijing Jiaotong University, Beijing 100044, China, and also with the School of Electronic Engineering and Computer Science, Queen Mary University of London, London E1 4NS, U.K. (email: twhou@bjtu.edu.cn). }
\thanks{Anna Li is with the School of Computing and Communications, Lancaster University, Lancaster LA1 4WA, U.K. (e-mail: a.li16@lancaster.ac.uk). }
\thanks{Sofie Pollin is with the  Department of Electrical Engineering (ESAT), KU Leuve, Belgium (e-mail: sofie.pollin@esat.kuleuven.be). }
\thanks{Yuanwei Liu is with the Department of Electrical and Electronic Engineering, The University of Hong Kong, Hong Kong (e-mail: yuanwei@hku.hk). }
\thanks{Arumugam Nallanathan is with the School of Electronic Engineering and Computer Science, Queen Mary University of London, London E1 4NS, U.K., and also with the Department of Electronic Engineering, Kyung Hee University, Yongin-si, Gyeonggi-do 17104, Korea (e-mail: a.nallanathan@qmul.ac.uk).}}

\maketitle

\begin{abstract}
With the rapid development of satellite communication and navigation, there is an urgent need to integrate both technologies to achieve reliable communication and precise navigation services within the same satellite system. By combining multi-/uni-cast (MUC) and non-orthogonal multiple access (NOMA) technologies, we propose a novel MUC-NOMA-based integrated navigation and communication (INAC) signal structure, in which the navigation and communication signals share a common pseudo noise (PN) sequence, thereby integrating satellite communication and navigation at the signal level. According to different power allocation strategies, two scenarios are defined: multi-cast-oriented (MO-) INAC and uni-cast-oriented (UO-) INAC, where a greater portion of power is assigned to either the multi-cast or the uni-cast signal, respectively. To mitigate co-channel interference, we employ successive interference cancellation (SIC) at the receiver and design a signal processing algorithm for the proposed INAC signal. Then, closed-form expressions are subsequently derived for the bit error rates (BER) of both the navigation and communication signals, along with the positioning accuracy of the navigation signal. To gain further insights, the impacts of power allocation factors and communication rates are evaluated. Our analysis results show that: i) In the MO-INAC scenario, the positioning and BER performance of navigation signal are excellent when more power is assigned to the multi-cast signal; ii) In the UO-INAC scenario, interference in the shared resources is reduced when more power is assigned to the uni-cast signal; iii) The ranging accuracy decreases as the communication data rate increases. Numerical results confirm the superior BER and positioning accuracy of the MO-INAC scenario for MEO satellites.
\end{abstract}

\begin{IEEEkeywords}
BER, INAC, positioning, MUC-NOMA, SIC
\end{IEEEkeywords}

\section{Introduction}
The global navigation satellite system (GNSS) is widely recognized for providing accurate positioning and timing services across all weather conditions~\cite{1}, with extensive applications in critical global industries including transportation, telecommunications, and agriculture~\cite{2},~\cite{3},~\cite{4}. Over decades of development, major GNSS implementations such as the global positioning system, Galileo, GLONASS, and the beidou satellite navigation system have become operational~\cite{5}. In parallel, satellite communications have emerged as indispensable infrastructure, particularly for assisting terrestrial networks and enabling emergency communications in remote or harsh environments such as oceans, mountainous regions, and deserts~\cite{6},~\cite{7}. In addition, satellite communications further support television broadcasting, maritime operations, and military activities by leveraging their inherent advantages of global coverage and exceptional reliability\cite{8}. Despite their complementary functionalities, traditional satellite navigation and communication systems face inherent inefficiencies. Traditional satellite navigation and communication rely on different orbital and spectrum resources, as well as other non-renewable resources, resulting in large power loss and low spectrum efficiency\cite{9},~\cite{10}. To address the challenge of sustainable development, researchers have actively investigated integrated navigation and communication (INAC) scheme. Liu et al. described three evolutionary stages of low earth orbit (LEO)-INAC, which introduce an innovative LEO-INAC architecture to enhance high-speed communication and high-precision navigation for emerging applications~\cite{new1}. Wang et al. explored the rationale and practicality of combining communication and navigation on LEO satellite platforms, evaluating different LEO-based navigation models and their respective strengths and weaknesses~\cite{new2}. However, the specific signal structure for INAC was not addressed in these studies.

The INAC signal refers to the signal that combines both navigation and communication functions into a single transmission, enabling the simultaneous provision of positioning, timing, and communication services, thereby optimizing the use of available resources and improving system efficiency~\cite{11},~\cite{12}. Recently, the researchers mainly employ time division, frequency division or code division multiplexing to fuse communication and navigation signals. For the INAC system, Wang et al. proposed a spreading code based on the Kasami sequence, comparing the performance with m-sequence in terms of correlation, balance, and the number of code groups~\cite{13}. Ma et al. designed a cyclic code shift keying (CCSK) based INAC signal~\cite{14}. The communication signal is modulated with pseudo-random noise (PN) sequence as the navigation signal using CCSK. To avoid cross-correlation interference between the navigation and communication signals, Zou et al. proposed a code-phase optimization algorithm, which shifts the peak of cross-correlation function to align with that of the autocorrelation function, effectively mitigating interference from cross-correlation and improving communication reliability~\cite{15}. To address the issue of multipath effects, a master-slave reception architecture employing a Rake receiver was also developed~\cite{16}. Feng et al. investigated the vector orthogonal frequency division multiplexing (OFDM) technique, which transmitted communication signals and pilot signals on different sub-carriers~\cite{17}. Zhou et al. proposed a multi-carrier direct sequence code-division multiple access modulation method based on the OFDM system, where the positioning signal was superimposed with the OFDM modulated communication signal~\cite{18}. Aiming at the challenge of anti-interference and stable operation in navigation denial scenario, Ma et al. designed an INAC waveform for distributed swarm systems~\cite{19}. Deng et al. introduced a time-division code-division OFDM positioning signal, which embedded a low-power positioning signal in the background noise of the communication signal, allowing continuous capture and tracking of the positioning signal~\cite{20},~\cite{21}. Huang et al. utilized the frequency diverse array (FDA) and OFDM techniques to develop the FDA-OFDM scheme for integrated communication, navigation and sensing applications~\cite{22}. Four INAC signal structures were proposed by Cui et al.:e.g., I/Q branch multiplexing, time-division multiplexing, frequency-division multiplexing, and code-division multiplexing~\cite{23}.

Although the studies mentioned above successfully fuse navigation and communication signals, time-division multiplexed INAC signal structure suffers from the discontinuity of the navigation signal, which hinders signal capture and tracking, thereby degrading ranging performance. In frequency-division multiplexing structure, both the navigation and communication signals occupy a substantial portion of the frequency bandwidth. In code-division multiplexing structure, the number of users is constrained by the limited number of spreading sequence. Consequently, the traditional INAC signal structures cannot achieve effective resource utilization, nor can it meet actual communication and navigation needs.

In order to provide reliable communication and high-precision navigation integrated services, the navigation signal has to be transmitted in continuous time resources. Non-orthogonal multiple access (NOMA), as a highly effective and advanced technique for optimizing resource utilization and enhancing system capacity, has been proposed~\cite{24,25,26,27}. Zhao et al. proposed a NOMA framework for internet of things networks, which achieves significantly higher system sum rate and channel gains than benchmark schemes~\cite{NOMA-new2}. Hou et al. proposed an innovative NOMA-INAC network by utilizing medium earth orbit (MEO) satellites~\cite{28},~\cite{29},~\cite{new3}. Yin et al. and Wang et al. have investigated the non-orthogonal superposition of navigation and communication signals, which is similar in concept to NOMA~\cite{30},~\cite{31}. However, both~\cite{30} and~\cite{31} only proposed the INAC signal, without providing practical solutions for interference elimination. In the case of NOMA, the receiver employs successive interference cancellation (SIC) to mitigate co-channel interference, thereby improving performance~\cite{32}. This drives us to design a NOMA-inspired interference cancellation scheme for the INAC signal at the receiver. While theoretically scalable, the practical applications of NOMA are often limited to two users per resource block. This is because the performance gains diminish rapidly with an increasing number of superimposed users due to the challenges of escalating receiver complexity and the detrimental effects of SIC error propagation~\cite{NOMA-new}.

\subsection{Motivation and Contribution}
Among the papers mentioned above, it is clear that existing studies have mainly concentrated on time-division, frequency-division and code-division multiplexed INAC signal structures, which suffer from poor positioning performance and low resource utilization. Although the NOMA-INAC signal has been partially addressed in~\cite{30} and~\cite{31}, a comprehensive performance analysis of INAC signal remains largely unexplored. The primary motivations for our work can be summarized as follows:
\begin{itemize}
  \item Since the navigation signal should be continuous, the NOMA technique offers a more practical solution than time-division and frequency-division multiplexing.
  \item Traditional performance analyses of INAC signals are primarily conducted from a channel perspective, whereas signal-level analyses, including BER and positioning accuracy, remain unexplored.
  \item The performance differences between  the low-power navigation signal scenario and the low-power communication signal scenario have not yet been determined.
\end{itemize}

To address these gaps, in this paper, we design a multi-/uni-cast (MUC) NOMA-INAC signal.  We compose the navigation signal with a portion of the communication signal to form a multi-cast signal, while the remaining communication signal is transmitted as a uni-cast signal. Then, we propose two potential power allocation scenarios, namely multi-cast-oriented INAC (MO-INAC) and uni-cast-oriented INAC (UO-INAC) scenarios.

The main contributions of this paper are summarized as follows:
\begin{itemize}
\item We propose a MUC-NOMA-based INAC signal architecture, where the low-rate multi-cast and the high-rate uni-cast signals are superimposed over the same PN sequence. Unlike conventional INAC systems where two services are typically allocated orthogonal resources, the proposed design enables navigation and communication to share the same time, frequency, and code resources within a direct sequence spread spectrum (DSSS) framework at the bit level. Based on the power allocation between the multi-cast and uni-cast components, two representative operating scenarios, MO-INAC and UO-INAC, are defined. Then, we design the signal demodulation and SIC detection processes, which provide a comprehensive framework for both MO-INAC and UO-INAC scenarios.

\item We develop a unified theoretical framework to analyze the BER and ranging performance of the proposed INAC system. Unlike conventional NOMA systems that typically assume identical symbol rates over narrow-band or OFDM channels, our proposed rate-split DSSS structure fundamentally changes the interference behavior at the receiver, leading to distinct error propagation mechanisms for the MO-INAC and UO-INAC scenarios. To characterize these effects, we develop a unified BER analytical framework, including the derivation of the despread decision statistics, the corresponding probability density functions (PDF), and the closed-form BER expressions for multi-cast and uni-cast signals. By incorporating the flexible resource splitting parameter, we derive exact closed-form BER expressions for navigation and communication signals in both MO-INAC and UO-INAC scenarios. Furthermore, we analytically map the residual interference to the delay-locked loop tracking jitter, providing closed-form expressions for the ranging accuracy.

\item To gain deeper insights, we derive the approximate closed-form BER expressions for multi-cast and uni-cast signals in both MO-INAC and UO-INAC scenarios under the high signal-to-noise ratio (SNR) condition, which explicitly reveal the fundamental relationships between the system performance and key design parameters. Our analysis demonstrates that the BER and positioning performance depend on the power allocation factors and communication rates. Furthermore, we reveal that there is a trade-off between the BER of the communication signal and that of the navigation signal.  

\item The simulation results confirm our analysis, demonstrating that: 1) In the MO-INAC scenario, the BER performance difference between the navigation and communication signals grows as the navigation signal power allocation factor or communication rate increases. 2) In the UO-INAC scenario, both the communication and navigation signals exhibit good BER performance when the power allocation factor for the communication signal is high enough and the communication rate is low enough. 3) In both MO-INAC and UO-INAC scenarios, higher communication data rates lead to reduced ranging accuracy. 4) Comprehensive baseline comparisons demonstrate the advantages of our proposed scheme in terms of communication reliability and resource conservation.
\end{itemize}

\subsection{Organization and Notations}

The structure of this paper is as follows. Section \uppercase\expandafter{\romannumeral2} outlines the model for the MUC-NOMA-based INAC signal. Section \uppercase\expandafter{\romannumeral3} describes the signal processing algorithms. In Section \uppercase\expandafter{\romannumeral4}, analytical derivations for the MO-INAC and UO-INAC scenarios are presented. Section \uppercase\expandafter{\romannumeral5} discusses the numerical evaluations of the proposed INAC signal. The final section summarizes the main conclusions of our work. Throughout the paper, the notation ${P_r}\left( \cdot\right)$ and $\mathbb{E}(\cdot)$ represent probability and expectation, respectively.

\section{System Model}

\begin{figure}[t!]
\centering
\includegraphics[width =3.5in]{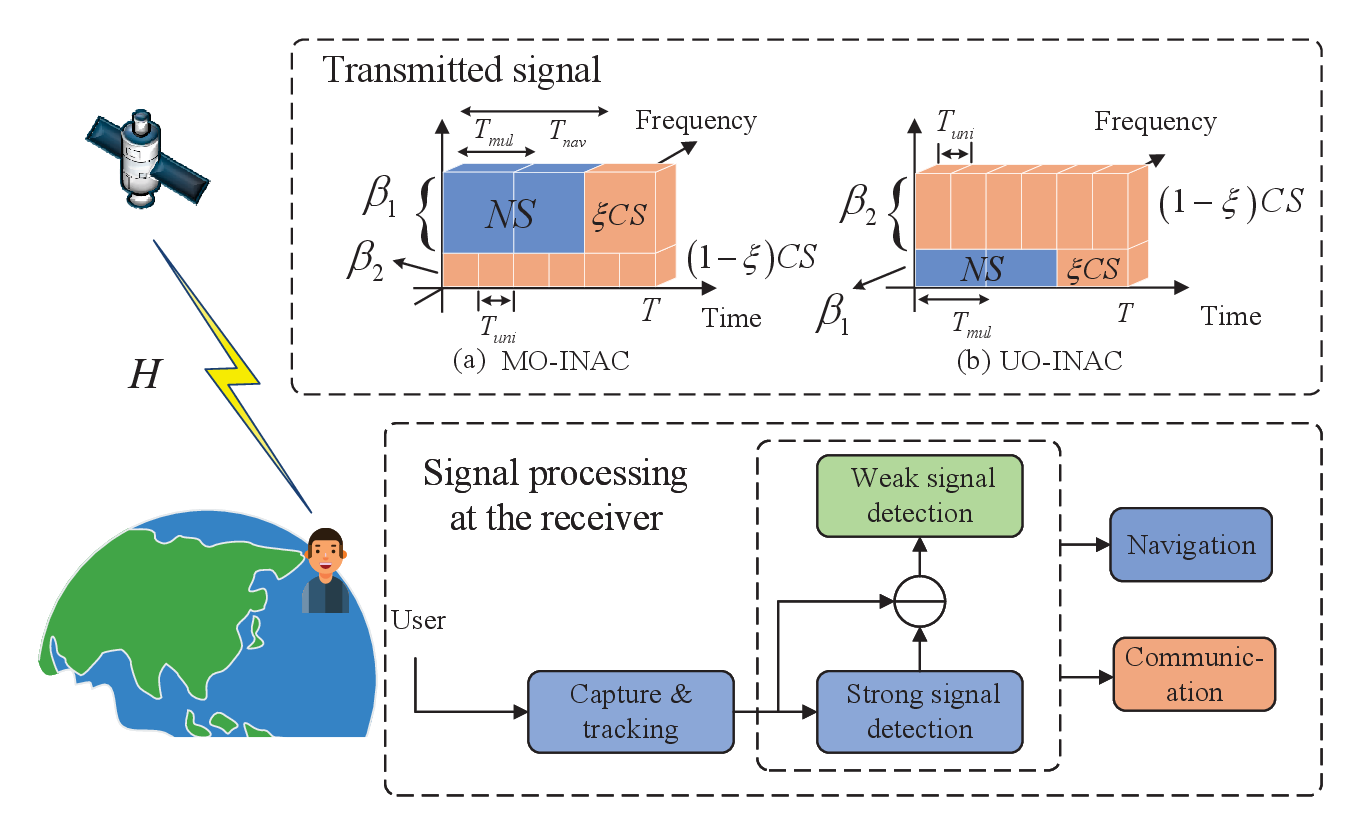}
\caption{MUC-NOMA-based INAC system model.}
\label{signal_model}
\end{figure}

In this paper, we examine a MUC-NOMA-based INAC signal. The system model of the proposed MUC-NOMA-based INAC is illustrated in Fig.~\ref{signal_model}, where both the satellite and the user are equipped with a single antenna. As shown in Fig.~\ref{signal_model}, the transmitted signal is organized into resource blocks in both time, frequency and power domains. Here, $NS$ and $CS$ denote the navigation signal and communication signal, respectively, while $\xi$ represents the fraction of communication resources allocated to the multi-cast service. On the one hand, when the interference power is strong, $\xi$ is set to 0, and $NS$ and $CS$ are directly superimposed by NOMA technique. On the other hand, when the interference power falls below a certain threshold, $\xi$ can be set to any value between 0 and 1. In this case, the communication signal $CS$ is split into $\xi CS$ and $(1-\xi)CS$. Then, $NS$ and $\xi CS$ are merged into a multi-cast signal, while $(1-\xi)CS$ forms a uni-cast signal. Then, the multi-cast and uni-cast signals are superimposed by NOMA technique.

The power allocation factors $\beta_1$ and $\beta_2$, which satisfy $\beta_1 + \beta_2 = 1$, are applied to the multi-cast and uni-cast signals, respectively. Based on these factors, the INAC signals are categorized into two scenarios: MO-INAC and UO-INAC. In the MO-INAC scenario, the upper block and lower block represent the multi-cast signal and uni-cast signal, respectively, with more power assigned to the multi-cast signal, i.e., $\beta_1 > \beta_2$. In the UO-INAC scenario, the upper block represents the uni-cast signal, whereas the lower block denotes the multi-cast signal. Here, more power is allocated to the uni-cast signal, i.e., $\beta_1 < \beta_2$. $T_{mul}$ and $T_{uni}$ are the multi-cast and uni-cast symbol periods, respectively. $T$ is the INAC signal duration, and $T_{nav}$ is the navigation signal duration. Note that $T_{uni} < T_{mul} $ because the uni-cast signal has higher data rate than the multi-cast signal, resulting in a shorter symbol period for uni-cast symbol.

\subsection{Transmitted Signal Model}

In this paper, since low-earth-orbit satellites cannot provide sufficient time dilution of precision and position dilution of precision for navigation, whereas high-earth-orbit satellites suffer from significant losses and substantial doppler delays, we focus on MEO satellites for providing both navigation and communication services to ground users~\cite{28}. In downlink transmission, inspired by the NOMA technique, navigation and communication signals can be composited to increase the channel capacity. We first define the discrete navigation symbols as ${n_v}\left[ k \right]$, split communication symbols as $a\left[ m \right]$ and uni-cast symbols as $b\left[ n \right]$. The split communication symbols $a\left[ m \right]$ correspond to the $\xi CS$ portion of the communication service shown in Fig.~\ref{signal_model}, while the uni-cast symbols $b\left[ n \right]$ correspond to the $(1-\xi)CS$ portion. Similar to satellite navigation, we adopt the binary phase shift keying technique~\cite{33}. Thus, the discrete navigation symbols, split communication symbols and uni-cast symbols can be respectively written as:
\begin{equation}\label{symbol define}
\begin{aligned}
&{n_v}\left[ k \right] \in \left\{ { - 1,1} \right\},~~k = 1,2, \ldots ,{K_{nav}},\\
&a\left[ m \right] \in \left\{ { - 1,1} \right\},~~m = 1,2, \ldots ,{K_{mul}},\\
&b\left[ n \right] \in \left\{ { - 1,1} \right\},~~~n = 1,2, \ldots ,{K_{uni}},
\end{aligned}
\end{equation}
where ${K_{nav}} = {T_{nav}}/{T_{mul}}$, ${K_{mul}} = \left( {T - {T_{nav}}} \right)/{T_{mul}}$ and ${K_{uni}} = T/{T_{uni}}$ represent the number of navigation, split communication and uni-cast symbols, respectively.

Then, the multi-cast signal can be written as:
\begin{equation}\label{cascade signal}
\begin{aligned}
&m\left( t \right) = \sum\nolimits_{k = 1}^{{K_{nav}}} {{n_v}\left[ k \right]{g_{tm}}\left( {t - \left( {k - 1} \right){T_{mul}}} \right) } \\
&+ \sum\nolimits_{m = 1}^{{K_{mul}}} {a\left[ m \right]{g_{tm}}\left( {t - {T_{nav}} - \left( {m - 1} \right){T_{mul}}} \right)}, 
\end{aligned}
\end{equation}
where $g_{tm}$ is the pulse shaping function of multi-cast signal at transmitter.

Similarly, the uni-cast signal is given by:
\begin{equation}\label{uni-cast signal}
\begin{aligned}
u\left( t \right) = \sum\nolimits_{n = 1}^{{K_{uni}}} {b\left[ n \right]{g_{tu}}\left( {t - \left( {n - 1} \right){T_{uni}}} \right)},
\end{aligned}
\end{equation}
where $g_{tu}$ is the pulse shaping function of uni-cast signal at transmitter.

The multi-cast and uni-cast signals are modulated with the same PN sequence separately, obtaining spread spectrum signals. Then, the multi-cast and uni-cast spread spectrum signals are superimposed by NOMA technique, yielding the MUC-NOMA-based INAC signal, which can be written as:
\begin{equation}\label{superimposed symbols}
s(t) = \left( {\sqrt {{\beta _1}} m(t)p(t) + \sqrt {{\beta _2}} u(t)p(t) } \right),
\end{equation}
where $p(t)$ is the PN signal with chip rate $T_c$, which can be written as:
\begin{equation}\label{PN code}
\begin{aligned}
p\left( t \right) = \sum\nolimits_{k = 1}^{{K_p}} {c\left[ k \right]{g_{tp}}\left( {t - k{T_c}} \right)},
\end{aligned}
\end{equation}
where $K_p$ represents the length of PN sequence, and $c\left[ k \right]$ is the discrete PN sequence. ${g_{tp}}$ denotes the pulse shaping function of PN signal.

The MUC-NOMA-based INAC signal is modulated with carrier, obtaining transmitted signal, which can be expressed as:
\begin{equation}\label{transmitted signal}
{T_x}(t) = \sqrt {2P} s(t)\cos \left( {2\pi {f_c}t + \varphi } \right),
\end{equation}
where ${f_c}$ and ${\varphi }$ represent the carrier frequency and carrier phase, respectively.  $ P $ denotes the satellite power.

\subsection{Channel Model}

Consider using a composite channel model that consists of two parts, large-scale fading and small-scale fading. The large-scale fading channel between the satellite and ground-user link is defined as:
\begin{equation}\label{The large-scale fading channel}
h_{lar}\left( d \right) = {\frac{c}{{4\pi {f_c}}d}},
\end{equation}
where ${d}$ denotes the distance between the satellite and ground-user. $c $ represents the speed of light.

Since satellite navigation signals use right-hand circularly polarized waves, multipath signals become left-hand circularly polarized after an odd number of reflections and can thus be filtered out. On the other hand, multipath components undergoing an even number of reflections are significantly weakened, making their impact on the received signal negligible. Additionally, satellite navigation is typically used in outdoor environments with few reflective surfaces. Therefore, for the small-scale channel model, only the phase shift due to free-space propagation distance is taken into account, allowing for a simplified representation as:
\begin{equation}\label{simplified small-scale fading channel}
h_{sm}\left( d \right) = \exp \left( { - \frac{{j2\pi d{f_c}}}{c}} \right).
\end{equation}

The composite channel can be written as:
\begin{equation}\label{The composite channel}
h \left( d \right) = {\frac{c}{{4\pi {f_c}}d}} \exp \left( { - \frac{{j2\pi d{f_c}}}{c}} \right).
\end{equation}

\subsection{Received Signal Model}

The received signal at the downlink user can be modeled as:
\begin{equation}\label{Received Signal Model}
\begin{aligned}
{R_x}(t) &= h\left( d \right){T_x}(t - {\tau _0}) + N(t)\\
 &= \gamma \sqrt {2P} s(t - {\tau _0})\cos (2\pi {f_c}(t - {\tau _0}) + \varphi ) + N(t),
\end{aligned}
\end{equation}
where $\gamma  = |h\left( d \right)|$ is the channel gain, and ${\tau _0} > 0 $ is the signal propagation delay. $N(t)$ is the zero-mean additive white Gaussian noise (AWGN) with the variance ${\sigma_{n} ^2}$. Assuming perfect estimation and compensation of the channel phase at the receiver, we employ down-converting to the received signal by the carrier. Subsequently, it is assumed that the down-converting signal is processed by an ideal low-pass filter, then the received signal can be transformed into:
\begin{equation}\label{low-pass filtered output signal}
\begin{aligned}
r(t) = \sqrt {2{P_s}} s(t - {\tau _0}) + n(t),
\end{aligned}
\end{equation}
where $P_s$ denotes the signal power at the receiver, and $n(t)$ represents the filtered AWGN signal.

\begin{remark}\label{signal design benifit}
The traditional composite signal, which modulates both navigation and communication signals onto a single continuous PN sequence, leads to ineffective communication when the data rate is low. Conversely, when the data rate is sufficiently high, it reduces the coherent integration time, degrading positioning performance. In our model, the multi-cast signal operates at a low rate for robust tracking, while the uni-cast communication signal is superimposed with multi-cast signal using the NOMA technique, enabling efficient high-rate data transmission without disrupting navigation performance.
\end{remark}

\subsection{Quantitative Resource Accounting}

Although the signal model described above is based on a basic two-user building block, in which one multi-cast stream and one uni-cast stream share a single PN sequence, the associated system-level gain in reducing code-resource scarcity can still be quantified.

Consider a practical satellite scenario in which a continuous navigation signal is broadcast while $K$ dedicated communication users are simultaneously served. In a conventional code-division multiplexing system, one PN sequence is allocated to the navigation broadcast, and $K$ additional quasi-orthogonal PN sequences are required for the $K$ communication users, resulting in a total of $K+1$ codes. In the proposed architecture, the navigation stream and one communication stream share a single PN sequence through MUC-NOMA, such that one communication link can be supported without requiring an additional PN sequence. This immediately reduces the required number of PN sequences by one. More generally, under a pairwise-sharing extension of the proposed framework, the remaining $K-1$ communication users can also be grouped over shared sequences, requiring $\frac{K-1}{2}$ additional PN sequences. In this case, the total number of required PN sequences becomes $\frac{K+1}{2}$, corresponding to up to $50\%$ savings compared with the conventional scheme.

In GNSS, high-quality spreading-code families are mathematically finite and extremely scarce. Therefore, the signal model investigated in this paper provides a fundamental physical-layer building block for code-efficient INAC architectures. Furthermore, the proposed INAC approach is readily extendable to emerging LEO satellite systems, where massive constellations and dense user access make efficient code-resource utilization even more important.

\section{Signal Processing Algorithms}

This section outlines the SIC processing flow for the INAC signal at the receiver. Since the core procedures for MO-INAC and UO-INAC are similar, with the primary distinction being the decoding order determined by their respective power allocation strategies, we provide a unified description of the algorithm.

\textbf{Step 1: Decode the signal with higher power (SHP).} To optimally detect the transmitted symbols in the presence of AWGN, a matched filter is employed at the receiver for each symbol. For the spread spectrum signal defined in our model, the matched filter can be efficiently implemented as a correlator. The received signal $r(t)$ is correlated with the local PN sequence ${\hat p}(t)$ to despread the SHP, which can be written as:
\begin{equation}\label{despread CAO-INAC signal}
{D_H}\left( t \right) = \int\limits_{(k_H - 1)T_H}^{k_H T_H} {r(t)\hat p(t)dt},
\end{equation}
where $k_H$ and $N_H$ denote the index and length of the symbols with higher power, respectively, and $k_H = 1,2,...N_H$. $T_H$ denotes the period of the symbols with higher power. In the MO-INAC scenario, $T_H = T_{mul}$, while in the UO-INAC scenario, $T_H = T_{uni}$.

Then, we perform decision-making on the despread signal ${D_H}\left( t \right)$, and the decoded symbols can be expressed as:
\begin{equation}\label{decoded cascade symbols}
{ S_H}(t - {k_H}{T_H}) = \left\{ \begin{aligned}
&~~~1 ~~{D_H}\left( t \right) > {V_t},\\
&-1 ~~{D_H}\left( t \right) < -{V_t},
\end{aligned} \right.
\end{equation}
where ${V_t}$ is the decision threshold.

\textbf{Step 2: Reconstruct SHP and perform SIC.} The decoded symbols with higher power are used to reconstruct the SHP, the reconstructed SHP can be written as:
\begin{equation}\label{reconstructed cascade signal}
s_{rec}^{H}(t) = \sqrt {2 P_s {\beta}} {S_H}(t){\hat p}(t),
\end{equation}
where $\beta$ represents the power allocation factor of SHP. In the MO-INAC scenario, $\beta = \beta_1$, while in the UO-INAC scenario, $\beta = \beta_2$.
 
By subtracting the reconstructed SHP from the original received INAC signal, the signal with lower power (SLP) at the receiver is obtained, which can be expressed as:
\begin{equation}\label{reconstructed communication signal}
s_{L}(t) ={r(t) - s_{rec}^{H}(t)}.
\end{equation}

\textbf{Step 3: Decode the SLP.} SLP is subsequently correlated with the PN sequence, and thus the despread SLP is given by:
\begin{equation}\label{despread PC signal}
{D_{L}}\left( t \right) = \int\limits_{(k_L - 1){T_L}}^{k_L{T_L}} s_{L}(t){\hat p}(t)dt,
\end{equation}
where $k_L$ and $ N_L$ denote the index and length of the symbols with lower power, respectively, with $k_L = 1,2,...{N_L}$. ${T_L}$ denotes the period of the symbols with lower power. In the MO-INAC scenario, $T_L = T_{uni}$, while in the UO-INAC scenario, $T_L = T_{mul}$.

The decoded symbols with lower power after decision-making can be expressed as:
\begin{equation}\label{ decoded PC symbols}
{S_L}(t - {k_L}{T_L}) = \left\{ \begin{aligned}
&~~~1~~~{D_L}\left( t \right) > {V_t},\\
&-1~~~{D_L}\left( t \right) < -{V_t}.
\end{aligned} \right.
\end{equation}

\textbf{Step 4: Separate multi-cast signal and merge uni-cast signal.} In the final step, the multi-cast symbols are split into navigation symbols ${n_v}\left[ k \right]$ and communication symbols $a\left[ m \right]$. Then, the communication symbols ${a}\left[ m \right]$ are combined with the decoded uni-cast symbols ${b}\left[ n \right]$ to form the complete decoded communication symbols.

\section{BER and Positioning Performance Analysis}
In this section, we develop a unified theoretical framework to analyze the BER performance of the proposed INAC signal. Then, we analyse the positioning performance of the INAC signal.

\subsection{Unified BER Performance Analysis}

Unlike conventional NOMA systems where superimposed users often share identical symbol rates, our proposed MUC-NOMA-based INAC structure is a rate-split NOMA system. Assuming perfect signal acquisition, the decision threshold ${V_t}$ can be set to $0$. Let $g_{mul}$ and $g_{uni}$ denote the spread spectrum gains of the multi-cast signal and uni-cast signal, respectively. The generalized rate ratio parameter is defined as $M$, representing the number of uni-cast symbols superimposed within one multi-cast symbol, which can be written as:
\begin{equation}\label{rate ratio M}
M = \frac{{\left( {1 - \xi } \right){R_{com}}}}{{{R_{nav}} + \xi {R_{com}}}} = \frac{{{T_{mul}}}}{{{T_{uni}}}} = \frac{{{g_{mul}}}}{{{g_{uni}}}},
\end{equation} 
where ${R_{nav}}$ and ${R_{com}}$ denote the data rate of navigation and communication signals, respectively.

Since the multi-cast signal contains the navigation data and a fraction $\xi$ of the communication data, while the uni-cast signal contains the remaining $\left( {1 - \xi } \right)$ portion, the BERs for the navigation signal $P_{eN}$ and the communication signal $P_{ec}$ can be unified as:
\begin{equation}\label{nav BER}
P_{eN} = P_{e,mul},
\end{equation}
and
\begin{equation}\label{com BER}
P_{ec} = \xi P_{e,mul} + \left( {1 - \xi } \right) P_{e,uni},
\end{equation}
where $P_{e,mul}$ and $P_{e,uni}$ are the BERs of the multi-cast and uni-cast signals, respectively.

\subsubsection{Detection of Higher-Power Signal} 

The signal detection process is shown in Section~III. We assume that the PN sequence at the transmitter and receiver is perfectly synchronized, thus ${\hat p}(t) = p(t - \tau _0) $. The decoded signal in \eqref{despread CAO-INAC signal} can be derived as:
\begin{equation}\label{rewritten correlator integral value}
\begin{aligned}
&{D_H}\left( t \right) = \int\limits_{{T_H}} {\left( {\sqrt {2{P_s}} s(t - {\tau _0}) + n(t)} \right)\hat p(t)dt} \\
&= \int\limits_{{T_H}} {\left( {\sqrt {{\beta _1}} m(t - {\tau _0}) + \sqrt {{\beta _2}} u(t - {\tau _0})} \right){{\hat p}^2}(t)dt} \\
&~~~~+ \int\limits_{{T_H}} {n\left( t \right)\hat p(t)dt} = {Term}_1 + {Term}_2,
\end{aligned}
\end{equation}
where ${Term}_1$ represents the signal component containing both the desired signal and the superimposed interference, and ${Term}_2 = \int_{T_H} n(t)\hat{p}(t)dt$ represents the noise component. 

For ${Term}_2$, since $n(t)$ is an AWGN signal and $\hat p(t)$ is a deterministic signal, the multiplication result of $n(t)$ and $\hat p(t)$ still obeys the Gaussian distribution. Moreover, $n(t)$ and $\hat p(t)$ are independently distributed, and thus the expectation and variance of the multiplication result can be respectively given by:
\begin{equation}\label{expectation}
{\mathbb E}\left[ {n(t)\hat p(t)} \right] = {\mathbb E}\left[ {n(t)} \right]{\mathbb E}\left[ {\hat p(t)} \right] = 0,
\end{equation}
and
\begin{equation}\label{variance}
\begin{aligned}
\sigma _{np}^2 &= {\mathbb E}[{n^2}(t){\hat p^2}(t)] - {\left\{ {{\mathbb E}[n(t)\hat p(t)]} \right\}^2}\\
&= {\mathbb E}[{n^2}(t){\hat p^2}(t)] = {\mathbb E}[{n^2}(t)] = \sigma _n^2.
\end{aligned}
\end{equation}

The central limit theorem in~\cite{34} establishes that the aggregate of numerous independent and identically distributed random variables approaches a Gaussian distribution. The integral value of the multiplication result ${Term}_2$ in a signal period can be considered as the summation of sufficient Gaussian signals. Therefore, ${Term}_2$ still obeys the Gaussian distribution. Since the expectation and variance of a Gaussian distribution are additive, the expectation and variance of ${Term}_2$ are respectively given by:
\begin{equation}\label{expectation of the integral value}
{\mathbb E}[{Term}_2] = 0,
\end{equation}
and
\begin{equation}\label{variance of the integral value}
\sigma _{inp}^2 = \sigma _n^2 + \sigma _n^2 + ... + \sigma _n^2 = {g_{H}}\sigma _n^2,
\end{equation}
where $g_H$ is the spread spectrum gain of the higher-power signal.

The derivation of the signal component ${Term}_1$, however, diverges into two distinct mathematical paradigms based on the different data rate between the superimposed signals.

\textbf{i) MO-INAC scenario:} In the MO-INAC scenario, the multi-cast signal is decoded first. Since $T_{mul} > T_{uni}$, one low-rate multi-cast symbol contains $M$ independent high-rate uni-cast symbols. Let $i$ denote the number of uni-cast symbols with phase 0, where $i = 0, 1, \dots, M$. We define $Term_1$ as $A_i$. Without loss of generality, we fix the phase of the multi-cast symbol at 0, and the same analysis applies when the phase is $\pi$. When the phase of multi-cast and uni-cast symbols are 0, i.e., the multi-cast and uni-cast symbols are 1, the first term of (\ref{rewritten correlator integral value}) can be expressed as:
\begin{equation}\label{A1}
{A_M} = {g_{mul}}\left( {\sqrt {2P_s{\beta _1}}  + \sqrt {2P_s{\beta _2}} } \right).
\end{equation}

On the contrary, when the multi-cast symbol is 1 and the uni-cast symbols are -1, the first term of (\ref{rewritten correlator integral value}) can be expressed as:
\begin{equation}\label{A2}
{A_0} = {g_{mul}}\left( {\sqrt {2P_s{\beta _1}}  - \sqrt {2P_s{\beta _2}} } \right).
\end{equation}

Therefore, when the uni-cast symbols $ b\left[ k \right] \in \left\{ { - 1,1} \right\}$, the first term of (\ref{rewritten correlator integral value}) can be written as:
\begin{equation}\label{A}
A_i = \sqrt{{\frac{i}{M}{A_M^2} + \frac{{M - i}}{M}{A_0^2}}}.
\end{equation}

Then, the despread signal $D_H(t)$ follows Gaussian distributions centered at $\pm A_i$:
\begin{subequations}\label{the PDF of the despread CAO-INAC signal}
\begin{align}
    f_{m,i1}(y) &= \frac{1}{\sqrt{2\pi g_{mul}\sigma_n^2}} \exp\left(-\frac{(y - A_i)^2}{2g_{mul}\sigma_n^2}\right), \\
    f_{m,i2}(y) &= \frac{1}{\sqrt{2\pi g_{mul}\sigma_n^2}} \exp\left(-\frac{(y + A_i)^2}{2g_{mul}\sigma_n^2}\right).
\end{align}
\end{subequations}

\begin{theorem}\label{Theorem1:navigation signal BER expression}
\emph{In the MO-INAC scenario, where ${\beta _1} > {\beta _2}$, the closed-form BER expression of multi-cast signal can be written as:}
\begin{equation}\label{navigation signal BER expression in theorem1}
\begin{aligned}
P_{e,mul}^{MO} = \sum\limits_{i = 0}^M {\frac{{C_M^i}}{{4\sum\limits_{i = 0}^M {C_M^i} }}} {\rm{erfc}}\left( {\sqrt {\frac{{{A_i}^2}}{{2{g_{mul}}\sigma _n^2}}} } \right).
\end{aligned}
\end{equation}

\begin{proof}
Please refer to Appendix A.
\end{proof}
\end{theorem}

\textbf{ii) UO-INAC Scenario:} In the UO-INAC scenario, the uni-cast signal is decoded first. Since $T_{uni} < T_{mul}$, the polarity of the interfering multi-cast symbol remains constant during one uni-cast symbol period. Therefore, ${Term}_1$ takes only two possible values. We define ${Term}_1$ as $A_{c,j}$, where $j$ denotes the multi-cast symbol with $j=1,-1$. To simplify the analysis, we assume that the uni-cast symbol is fixed at 1. When the multi-cast symbol is 1, ${A_{c,1}}$ is given by:
\begin{equation}\label{1-the first term in (16)}
{A_{c,1}} = {g_{uni}}\left( {\sqrt {2P_s{\beta _2}}  + \sqrt {2P_s{\beta _1}} } \right).
\end{equation}

On the contrary, when the multi-cast symbol is -1, ${A_{c,-1}}$ is given by:
\begin{equation}\label{2-the first term in (16)}
{A_{c,-1}} = {g_{uni}}\left( {\sqrt {2P_s{\beta _2}}  - \sqrt {2P_s{\beta _1}} } \right).
\end{equation}

The probability for $A_{c,j}$ is $1/2$. Similar to the PDFs in \eqref{the PDF of the despread CAO-INAC signal}, the corresponding PDFs of the despread signal in UO-INAC scenario are Gaussian distributions centered at $\pm A_{c,j}$ with variance $g_{uni}\sigma_n^2$.

\begin{theorem}\label{Theorem3:CO-BER expression of communication signal}
\emph{In the UO-INAC scenario, where ${\beta _1} < {\beta _2}$, the closed-form BER expression of uni-cast signal can be written as:}
\begin{equation}\label{CO-communication signal BER expression in theorem3}
\begin{aligned}
P_{e,uni}^{UO} = \frac{1}{4}{\rm{erfc}}\left( {\sqrt {\frac{{A_{c,1}^2}}{{2{g_{uni}}\sigma _n^2}}} } \right) + \frac{1}{4}{\rm{erfc}}\left( {\sqrt {\frac{{A_{c, - 1}^2}}{{2{g_{uni}}\sigma _n^2}}} } \right).
\end{aligned}
\end{equation}

\begin{proof}
The proof is similar to that in Appendix A.
\end{proof}
\end{theorem}

However, it is challenging to directly extract engineering insights from \eqref{navigation signal BER expression in theorem1} and \eqref{CO-communication signal BER expression in theorem3}. Therefore, to gain further insights, we investigate the approximate expression under high SNR conditions.

\begin{corollary}\label{corollary1:approximated CAO_navigation BER}
\emph{In the high SNR regime, assuming ${\beta _1} > {\beta _2}$, the BER of the multi-cast signal in the MO-INAC scenario can be approximated in closed-form as:}
\begin{equation}\label{Corro1:approximated CAO_navigation BER}
\begin{aligned}
\overline {P_{e,mul}^{MO}}  &\approx \frac{1}{{{2^M}}}\left[ {\frac{1}{6}\exp \left( { - \gamma {{\left( {\sqrt {{\beta _1}}  - \sqrt {{\beta _2}} } \right)}^2}} \right)} \right.\\
&\left. { + \frac{1}{2}\exp \left( { - \frac{4}{3}\gamma {{\left( {\sqrt {{\beta _1}}  - \sqrt {{\beta _2}} } \right)}^2}} \right)} \right],
\end{aligned}
\end{equation}
\emph{where $\gamma  = {P_s}{g_{mul}}/\sigma _n^2$. }
\begin{proof}
Please refer to Appendix B.
\end{proof}
\end{corollary}

\begin{remark}\label{rate ratio CAO_INAC remark1}
From \eqref{Corro1:approximated CAO_navigation BER}, we can see that the BER is proportional to the probability factor $1/2^M$. As the rate ratio parameter $M$ increases, the BER of the multi-cast signal decreases significantly.
\end{remark}

\begin{remark}\label{beta CAO_INAC remark2}
Since the power allocation factors satisfy ${\beta _1} > {\beta _2}$ and ${\beta _1} + {\beta _2} = 1$, increasing the multi-cast power allocation factor ${\beta _1}$ directly enlarges the effective amplitude difference $\left(\sqrt{\beta_1} - \sqrt{\beta_2}\right)$, which leads to an exponential reduction in the BER of the multi-cast signal.
\end{remark}

\begin{corollary}\label{corollary3:approximated CO communication BER}
\emph{When ${\beta _1} < {\beta _2}$, the BER of uni-cast signal in the UO-INAC scenario can be approximated in closed-form as:}
\begin{equation}\label{Corro3:approximated CO communication BER}
\begin{aligned}
\overline {P_{e,uni}^{UO}}  &= \left[ {\frac{1}{{24}}{e^{ - \frac{{{g_{uni}}{P_s}}}{{\sigma _n^2}}}} + \frac{1}{8}{e^{ - \frac{{4{g_{uni}}{P_s}}}{{3\sigma _n^2}}}}} \right]\\
 &\times \left[ {{e^{ - {{\left( {\sqrt {{\beta _1}}  + \sqrt {{\beta _2}} } \right)}^2}}} + {e^{ - {{\left( {\sqrt {{\beta _1}}  - \sqrt {{\beta _2}} } \right)}^2}}}} \right].
\end{aligned}
\end{equation}
\begin{proof}
By substituting the BER of communication signal in \eqref{CO-communication signal BER expression in theorem3} into error function \eqref{simple but good approximation for erfc(x)}, we obtain the BER expression in \eqref{Corro3:approximated CO communication BER}. The proof is complete.
\end{proof}
\end{corollary}

\begin{remark}\label{remark3}
The result of \eqref{Corro3:approximated CO communication BER} demonstrates that the BER of uni-cast signal in UO-INAC scenario increases as the spread spectrum gain of uni-cast signal $g_{uni}$ decreases.
\end{remark}

\begin{remark}\label{beta CO remark4}
Under the given constraints, a larger gap between $\beta_1$ and $\beta_2$ leads to a lower BER for the uni-cast signal in the UO-INAC scenario.
\end{remark}

\subsubsection{Detection of the Lower-Power Signal via SIC}

The BER of the lower-power signal $P_{r,L}$ depends on whether the first stage is decoded correctly or incorrectly. The unified BER expression of lower-power signal can be written as:
\begin{equation}\label{BER low power}
P_{r,L} = (1 - P_{r,H}) P_{L|C} + P_{r,H} P_{L|E},
\end{equation}
where $P_{r,H}$ is the BER of the higher-power signal obtained from the first stage, $P_{L|C}$ and $P_{L|E}$ represent the BER of the lower-power signal when the first stage was decoded correctly and incorrectly, respectively.

Because of the different data rate between the multi-cast and uni-cast signals, the error propagation mechanism exhibits fundamentally different characteristics in the two scenarios. In the following, we analyze them in turn.

\textbf{i) MO-INAC Scenario:} In this scenario, since the low-rate multi-cast signal is decoded first, for each high-rate uni-cast symbol, the residual interference from the first stage is deterministic and binary. 

\textbf{Case 1:} When the multi-cast symbol is correctly decoded, the residual signal is interference-free, and the despread uni-cast signal can be written as:
\begin{equation}\label{rewritten correlator integral value2 case1}
\begin{aligned}
{D_L}\left( t \right) = \int_{{T_{uni}}} {\sqrt {2{P_s}{\beta _2}} u(t - {\tau _0}){{\hat p}^2}(t) + n(t)\hat p(t)dt}.
\end{aligned}
\end{equation}

Then, the amplitude of the despread uni-cast signal is given by $A_{c,1}^{'} = g_{uni}\sqrt{2P_s\beta_2}$, and the BER of uni-cast symbols in \textbf{Case 1} is given by:
\begin{equation}\label{communication signal BER expression}
P_{e,uni|C} = \frac{1}{2}{\rm{erfc}}\left( {\sqrt {\frac{{{P_s}{\beta _2}{g_{uni}}}}{{\sigma _n^2}}} } \right).
\end{equation}

\textbf{Case 2:} When the multi-cast symbol is incorrectly decoded, the residual interference doubles. The despread uni-cast signal can be written as:
\begin{equation}\label{rewritten correlator integral value2 case2}
\begin{aligned}
&{D_L}\left( t \right) = 2\int_{{T_{uni}}} {\sqrt {2{P_s}{\beta _1}} m(t - {\tau _0}){{\hat p}^2}(t)dt} \\
&+ \int_{{T_{uni}}} {\sqrt {2{P_s}{\beta _2}} u(t - {\tau _0}){{\hat p}^2}(t) + n(t)\hat p(t)dt}.
\end{aligned}
\end{equation}

If the transmitted multi-cast and uni-cast symbols have the same polarity, the amplitude of the despread signal becomes $A_{c,2}^{'} = g_{uni}\sqrt{2P_s}(2\sqrt{\beta_1}+\sqrt{\beta_2})$. On the contrary, if the polarities of the transmit multi-cast and uni-cast symbols are different, the uni-cast symbol is flipped because ${\beta _1}>{\beta _2}$, and thus the uni-cast symbol will definitely be decoded incorrectly. Therefore, assuming that the multi-cast and uni-cast signals have an equal probability to transmit identical symbols, the BER of uni-cast symbols in \textbf{Case 2} can be given by:
\begin{equation}\label{communication signal BER expression in case 2}
\begin{aligned}
{P_{e,uni{\rm{|}}E}} = \frac{1}{2}{\rm{erfc}}\left( {\frac{{\sqrt {{P_s}{g_{uni}}} \left( {2\sqrt {{\beta _1}}  + \sqrt {{\beta _2}} } \right)}}{ {\sigma _n}}} \right) + \frac{1}{2}.
\end{aligned}
\end{equation}

\begin{theorem}\label{Theorem2:communication signal BER expression}
\emph{In the MO-INAC scenario, the closed-form BER expression of uni-cast signal averaged over all possible combinations can be written as:}
\begin{equation}\label{communication signal BER expression in theorem2}
\begin{aligned}
&P_{e,uni}^{MO} = \frac{1}{2}\left( {1 - P_{e,mul}^{MO}} \right){\rm{erfc}}\left( {\sqrt {\frac{{{P_s}{\beta _2}{g_{uni}}}}{{\sigma _n^2}}} } \right) \\
&+ \frac{1}{2}P_{e,mul}^{MO}\left[ {{\rm{erfc}}\left( {\frac{{\sqrt {{P_s}{g_{uni}}} \left( {2\sqrt {{\beta _1}}  + \sqrt {{\beta _2}} } \right)}}{{{\sigma _n}}}} \right) + 1} \right].
\end{aligned}
\end{equation}

\begin{proof}
By substituting the conditional BER in \eqref{communication signal BER expression} and \eqref{communication signal BER expression in case 2} into \eqref{BER low power}, the closed-form BER expression of uni-cast signal averaged over all possible combinations can be obtained. The proof is complete.
\end{proof}
\end{theorem}

\begin{corollary}\label{corollary2:approximated CAO_communication BER}
\emph{In the high SNR regime, it is obvious that the BER of communication signal mainly depends on the BER expression of uni-cast signal in \textbf{Case 1}. Then, the BER of the uni-cast signal in the MO-INAC scenario can be approximated by:}
\begin{equation}\label{Corro2:CAO_INAC_decoding communication}
\begin{aligned}
\overline {P_{e,uni}^{MO}}  = \frac{1}{{12}}{e^{ - \frac{{{P_s}{g_{uni}}{\beta _2}}}{{\sigma _n^2}}}} + \frac{1}{4}{e^{ - \frac{4}{3}\frac{{{P_s}{g_{uni}}{\beta _2}}}{{\sigma _n^2}}}}.
\end{aligned}
\end{equation}
\begin{proof}
In order to glean further engineering insights, the error function can be expanded similar to \eqref{simple but good approximation for erfc(x)}. We then substitute  the BER expression \eqref{communication signal BER expression} into the error function \eqref{simple but good approximation for erfc(x)}, obtaining the expression in \eqref{Corro2:CAO_INAC_decoding communication}. The proof is complete.
\end{proof}
\end{corollary}

\begin{remark}\label{remark 6} 
The result of ~\eqref{Corro2:CAO_INAC_decoding communication} proves that the BER of uni-cast signal decreases as the power allocation factor of uni-cast signal increases in the MO-INAC scenario.
\end{remark}

\textbf{ii) UO-INAC Scenario:} In the UO-INAC scenario, $M$ uni-cast symbols are first decoded and canceled within one low-rate multi-cast symbol period. If some of these $M$ symbols are decoded incorrectly, the resulting residual interference is accumulated from multiple random errors. Unlike the deterministic binary error in the MO-INAC scenario, this accumulated error behaves as random interference and degrades the effective SNR. Therefore, we introduce a BER impact factor $\delta \in (0,1)$ to characterize this degradation.

If the uni-cast symbols are correctly decoded, the residual signal is interference-free, which can be written as:
\begin{equation}\label{integral value shown in 20}
\begin{aligned}
{D_{L}}\left( t \right) = \int_{{T_{mul}}} {\sqrt {2{P_s}{\beta _1}} m(t - {\tau _0}){{\hat p}^2}(t)}  + n\left( t \right)\hat p(t)dt.
\end{aligned}
\end{equation}

Then, the amplitude of the multi-cast signal in UO-INAC scenario is given by $A_U = g_{mul}\sqrt{2P_s\beta_1}$. Considering the uncertain error propagation, the effective amplitude is scaled by $\sqrt{\delta}$.

\begin{theorem}\label{Theorem4:CO-BER expression of navigation signal}
\emph{In the UO-INAC scenario, where ${\beta _1} < {\beta _2}$, the closed-form BER expression of multi-cast signal can be written as:}
\begin{equation}\label{CO-navigation signal BER expression in theorem4}
\begin{aligned}
P_{e,mul}^{UO} = \frac{1}{2}{\rm{erfc}}\left( {\sqrt {\delta {P_s}{g_{mul}}{\beta _1}} /{\sigma _n}} \right).
\end{aligned}
\end{equation}
\begin{proof}
When the uni-cast symbols are correctly decoded, the BER is derived from the standard BPSK formulation over the AWGN channel similar to \eqref{communication signal BER expression}. Furthermore, the factor $\delta$ is introduced to capture the energy loss due to the uncertain accumulation of incorrect uni-cast symbol subtractions. The proof is complete.
\end{proof}
\end{theorem}

\begin{corollary}\label{corollary4:approximated CO navigation BER}
\emph{The BER of multi-cast signal in the UO-INAC scenario can be approximated in closed-form as:}
\begin{equation}\label{Corro4:approximated CO navigation BER}
\begin{aligned}
\overline {P_{e,mul}^{UO}}  = \frac{1}{{12}}{e^{ - \frac{{\delta {P_s}{g_{mul}}{\beta _1}}}{{\sigma _n^2}}}} + \frac{1}{4}{e^{ - \frac{4}{3}\frac{{\delta {P_s}{g_{mul}}{\beta _1}}}{{\sigma _n^2}}}}.
\end{aligned}
\end{equation}
\begin{proof}
By substituting the BER of the multi-cast signal in \eqref{CO-navigation signal BER expression in theorem4} into error function \eqref{simple but good approximation for erfc(x)}, we yield the BER expression in \eqref{Corro4:approximated CO navigation BER}. The proof is complete.
\end{proof}
\end{corollary}

\begin{remark}\label{delta CO remark7}
Due to the high communication data rates, if the uni-cast symbols are decoded incorrectly, the accumulated interference to the navigation signal will increase, significantly impacting the BER performance of the navigation signal. Therefore, the BER impact factor $\delta$ will become smaller as the rate factor $M$ increases.
\end{remark}

\subsubsection{BER Expressions for Navigation and Communication Signals}

Having derived the BERs for the multi-cast and the uni-cast streams in both scenarios, we now substitute them into the unified equations \eqref{nav BER} and \eqref{com BER} to obtain the BERs for the navigation and communication signals.

For the MO-INAC scenario, the BER of navigation signal is exactly the BER of multi-cast signal derived in Theorem \ref{Theorem1:navigation signal BER expression}, which is given by:
\begin{equation}\label{nav BER substitude}
    P_{eN}^{MO} = P_{e,mul}^{MO}.
\end{equation}
The overall BER of communication signal is the weighted sum of multi-cast and uni-cast BERs, which can be written as:
\begin{equation}\label{com BER substitude}
    P_{ec}^{MO} = \xi P_{e,mul}^{MO} + (1-\xi) P_{e,uni}^{MO}.
\end{equation}

For the UO-INAC scenario, the BER of navigation signal is given by the multi-cast BER derived in Theorem \ref{Theorem4:CO-BER expression of navigation signal}, which can be formulated as:
\begin{equation}
    P_{eN}^{UO} = P_{e,mul}^{UO}.
\end{equation}
The overall BER of the communication signal in the UO-INAC scenario is given by:
\begin{equation}
    P_{ec}^{UO} = \xi P_{e,mul}^{UO} + (1-\xi) P_{e,uni}^{UO}.
\end{equation}

\subsection{Ranging Performance Analysis}

In satellite navigation, ranging is often achieved through code phase tracking. Since the analysis results of SNR have already been provided in the derivation of the BER, the root mean square error of code phase measurement in the proposed MO-INAC scenario can be written as~\cite{code-phase-tracking}:
\begin{equation}\label{square error of code phase measurement-MO}
\begin{aligned}
\sigma _{tDLL}^{MO} = \sum\limits_{i = 0}^M {\frac{{C_M^i}}{{\sum\limits_{i = 0}^M {C_M^i} }}\sqrt {\frac{{{B_L}}}{{2\frac{{{P_{MO}}}}{{{N_0}}}}}\frac{1}{{{B_{fe}}{T_c}}}\left( {1 + \frac{1}{{{T_{coh}^{MO}}\frac{{{P_{MO}}}}{{{N_0}}}}}} \right)}},
\end{aligned}
\end{equation} 
where $\frac{{{P_{MO}}}}{{{N_0}}} = \frac{{{A_i}^2}}{{{g_{nav}}{N_0}}}$. $B_L$ and $B_{fe}$ represent the bandwidth of loop noise and transmitted signal, respectively, and $T_{coh}^{MO}$ denotes the coherent integration duration in the MO-INAC scenario. 

In the UO-INAC scenario, the strong communication signal is utilized for code phase tracking. The polarity reversal of communication data bit limits the coherent integration duration, leading to a trade-off between communication data rate and code phase tracking accuracy. For instance, a low data rate of 1 kbps allows for a long coherent integration duration of 1 ms, maximizing the processing gain, which is critical for ranging accuracy. Conversely, increasing the data rate to 100 kbps would reduce the maximum coherent integration duration to 10 µs, resulting in a 20 dB loss in processing gain. However, this trade-off does not limit the practical value of the UO-INAC scenario. In practice, the system can flexibly adapt the communication data rate: Employing higher rates in favorable channel conditions, and reverting to lower rates in weak-signal environments when reliable positioning is the priority. This adaptability is a key feature of the proposed INAC signal. The root mean square error of code phase measurement in the proposed UO-INAC scenario can be written as:
\begin{equation}\label{square error of code phase-UO}
\begin{aligned}
\sigma _{tDLL}^{UO} &= \frac{1}{2}\sqrt {\frac{{{B_L}}}{{2\frac{{A_{c,1}^2}}{{{g_{uni}}{N_0}}}}}\frac{1}{{{B_{fe}}{T_c}}}\left( {1 + \frac{1}{{{T_{coh}^{UO}}\frac{{A_{c,1}^2}}{{{g_{uni}}{N_0}}}}}} \right)} \\
 &+ \frac{1}{2}\sqrt {\frac{{{B_L}}}{{2\frac{{A_{c,1}^2}}{{{g_{uni}}{N_0}}}}}\frac{1}{{{B_{fe}}{T_c}}}\left( {1 + \frac{1}{{{T_{coh}^{UO}}\frac{{A_{c,1}^2}}{{{g_{uni}}{N_0}}}}}} \right)}, 
\end{aligned}
\end{equation}
where ${T_{coh}^{UO}}$ represents the correlation integration duration in the UO-INAC scenario.

Then, the ranging error of MO-INAC and UO-INAC scenarios can be respectively given by:
\begin{equation}\label{ranging error-MO}
{d_{e,MO}} = {c\sigma _{tDLL}^{MO}}{T_c},
\end{equation}
and
\begin{equation}\label{ranging error-UO}
{d_{e,UO}} = {c\sigma _{tDLL}^{UO}}{T_c}.
\end{equation}

\begin{remark}\label{remark-ranging}
Since the polarity reversal of communication data bit limits the coherent integration duration, the ranging accuracy decreases as the communication data rate increases.  
\end{remark}

\section{Numerical Studies}
This section provides numerical results for evaluating the performance of the MUC-NOMA-based INAC signal. Monte Carlo simulations with ${10^6}$ realizations are conducted to validate the accuracy of the analytical expressions. The system bandwidth is configured to 4.092 MHz, and the carrier frequency is set to 1207.14 MHz. The power of the AWGN is calculated as $\sigma _n^2 =  - 174 + 10{\rm{lo}}{{\rm{g}}_{10}}(B_{fe})$ dBm. The navigation data rate is set to 500 bit/s. The PN sequence is designed based on the spreading code of the BeiDou B1I navigation signal, with a chip rate of 2.046 Mchip/s and a sequence length of 2046 chips. Thus, the spread spectrum gain of navigation signal is set to ${g_{nav}} = 2046$. The distance $d$ between the satellite and the user ranges from 8000 km to 20000 km. In the MO-INAC scenario, the power allocation factors of the navigation and communication signals are set to ${\beta _1} = 0.70$ and ${\beta _2} = 0.30$, respectively, while in the UO-INAC scenario, these factors are set to ${\beta _1} = 0.30$ and ${\beta _2} = 0.70$, respectively.

\begin{table}[h!]
\centering
\caption{List of simulation parameters.}
\small {
\begin{tabular}{cc}
\toprule
\textbf{ Simulation parameter} & \textbf{ Value} \\
\midrule
 Bandwidth of signal $B_{fe}$ & 4.092 MHz \\
 Carrier frequency $f_c$ & 1207.14 MHz \\
 AWGN power & -137.86 dBW \\
 Navigation data rate & 500 bps \\
 Satellite-user distance $d$ & 8000-20000 km \\
 ${\beta _1}$:${\beta _2}$ in MO-INAC &  0.70:0.30 \\
 ${\beta _1}$:${\beta _2}$ in UO-INAC &  0.30:0.70 \\
 Chip rate $T_c$ & 1/2046 ms \\
 Spread spectrum gain ${g_{nav}} $ & 2046 \\
 Bandwidth of loop noise $B_L$        & 0.2 Hz \\
 Coherent integration duration $T_{coh}$    & 1 ms    \\
\bottomrule
\end{tabular}}
\label{simulation parameters}
\end{table}

\begin{figure}[t!]
\centering
\includegraphics[width =3.5in]{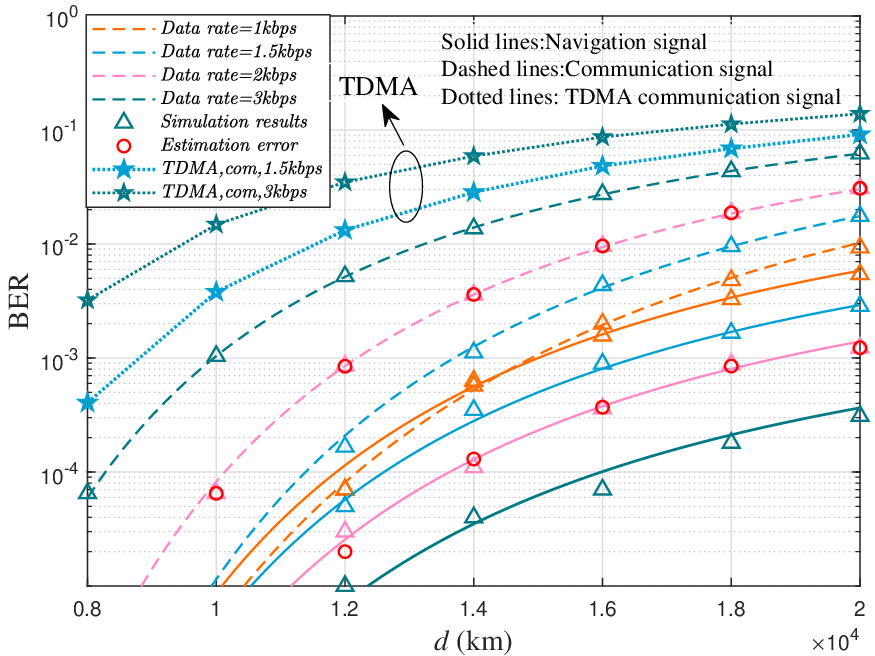}
\caption{The BER of MUC-NOMA-based INAC signal in MO-INAC scenario, where the analytical results are derived from (\ref{navigation signal BER expression in theorem1}) and (\ref{communication signal BER expression in theorem2}).}
\label{CAO-INAC-BER}
\end{figure}

\begin{figure}[t!]
\centering
\includegraphics[width =3.5in]{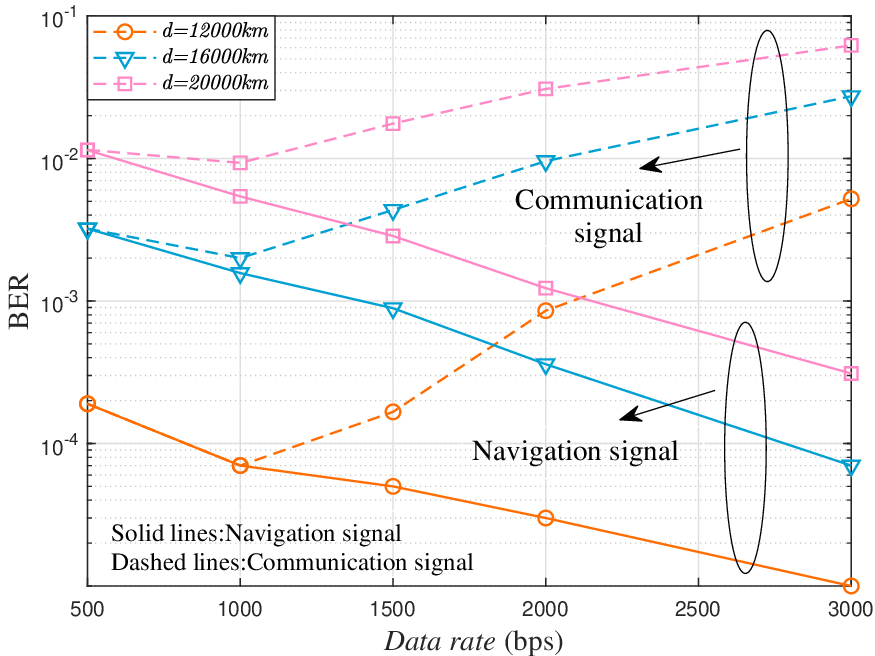}
\caption{The BER of the MO-INAC scenario against the communication data rate.}
\label{CAO-Rate_compare}
\end{figure}

\emph{1) Impact of the rate ratio parameter on the BER in the MO-INAC scenario:} Fig.~\ref{CAO-INAC-BER} depicts the BER of navigation and communication signals in the MO-INAC scenario. The theoretical results and Monte Carlo results of the navigation signal and communication signal are represented by solid lines, dashed lines, and triangles, respectively. It can be seen that the BER increases as the distance between the satellite and user increases. This is due to the fact that the path loss increases as distance increases. Moreover, we can observe that when communication data rate increases, the BER of navigation signal decreases, while the BER of communication signal increases, which validates our~\textbf{Remark~\ref{rate ratio CAO_INAC remark1}}. Monte Carlo simulations with practical satellite impairments such as residual Doppler shifts and phase noise after tracking loops are represented by circles. It can be observed that the simulation results with estimation errors show only a minor deviation from those with ideal channel estimation, indicating that the proposed system exhibits strong robustness and practical feasibility. Furthermore, as indicated by the dotted lines, the proposed NOMA scheme consistently outperforms the time-division multiple access (TDMA) baseline across different data rates, since TDMA halves the effective coherent integration time through orthogonal time partitioning. The TDMA navigation curves are omitted because their BER stays below $10^{-6}$ without intra-code interference, falling outside the displayed range.

In Fig.~\ref{CAO-Rate_compare}, we can also see that as the rate ratio parameter increases, that is, the number of communication symbols increases, the spread spectrum gain of the communication signal decreases significantly, resulting in an increased BER of the communication signal. The results also indicate that in the MO-INAC scenario, the BER performance of the navigation signal is better than that of the communication signal.

\begin{figure}[t!]
\centering
\includegraphics[width =3.5in]{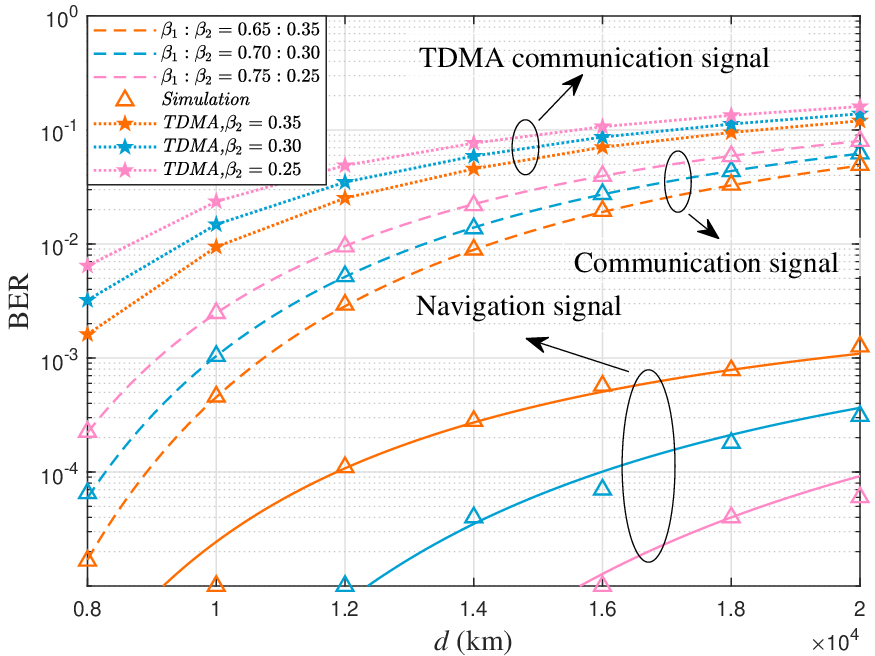}
\caption{The BER of the MO-INAC scenario under different power allocation factors. The communication data rate is set to 3 kbps.}
\label{CAO-beta_compare}
\end{figure}

\emph{2) Impact of the power allocation factor on BER in the MO-INAC scenario:} In Fig.~\ref{CAO-beta_compare}, we evaluate the BER in the MO-INAC scenario with different power allocation factors. We can see that as the power allocation factor of navigation signal increases, the BER of navigation signal decreases, and the BER of the communication signal increases, which validates our~\textbf{Remark~\ref{beta CAO_INAC remark2}} and~\textbf{Remark~\ref{remark 6}}. Notably, when the power allocation factor difference between the navigation and communication signals grows, the performance gap in BER between two signals also becomes more pronounced. This observation highlights a critical trade-off between the BER performance of the navigation and the communication signals, where optimizing one signal's performance comes at the cost of the other's. This effect is particularly evident in the curves with larger disparities in power allocation, such as $\beta_1$:$\beta_2$=0.75:0.25, where the navigation signal experiences the best performance at the cost of a considerable performance degradation in the communication signal. Additionally, the slope of the BER curves for the communication signal is noticeably higher compared to the navigation signal, especially at higher distances. This indicates that the communication signal is more sensitive to changes in the power allocation factor, which can be attributed to its higher reliance on the available power compared to the navigation signal. To further illustrate the advantage of the proposed scheme, the TDMA baseline under the corresponding power allocation factors is also included for comparison. The results show that the proposed NOMA scheme consistently outperforms the TDMA baseline, thereby demonstrating the robust communication performance of NOMA under limited power resources.

\begin{figure}[t!]
\centering
\includegraphics[width =3.5in]{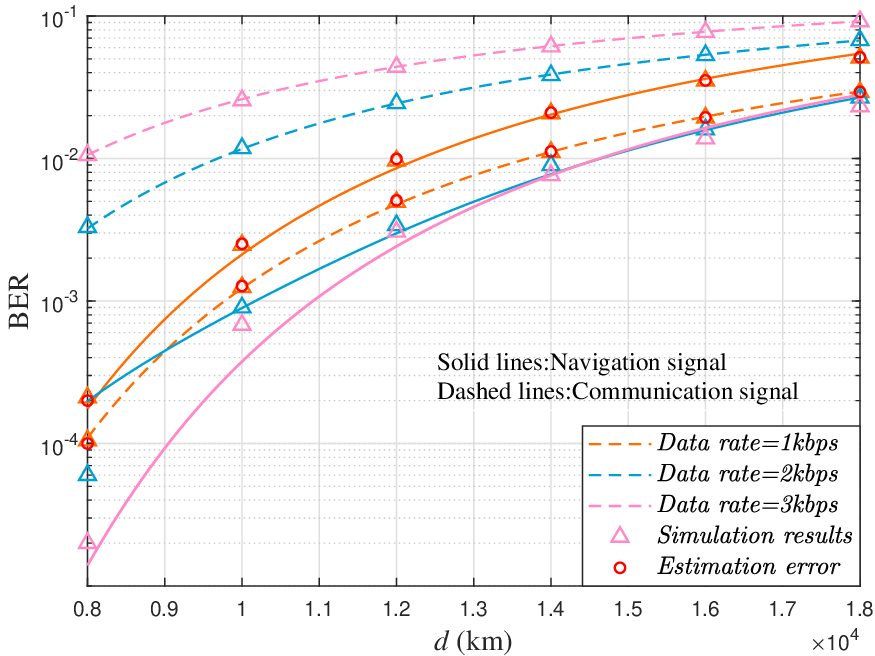}
\caption{The BER of MUC-NOMA-based INAC signal in the UO-INAC scenario, where the analytical results are derived from (\ref{CO-communication signal BER expression in theorem3}) and (\ref{CO-navigation signal BER expression in theorem4}).}
\label{CO-INAC-BER}
\end{figure}

\begin{figure}[t!]
\centering
\includegraphics[width =3.5in]{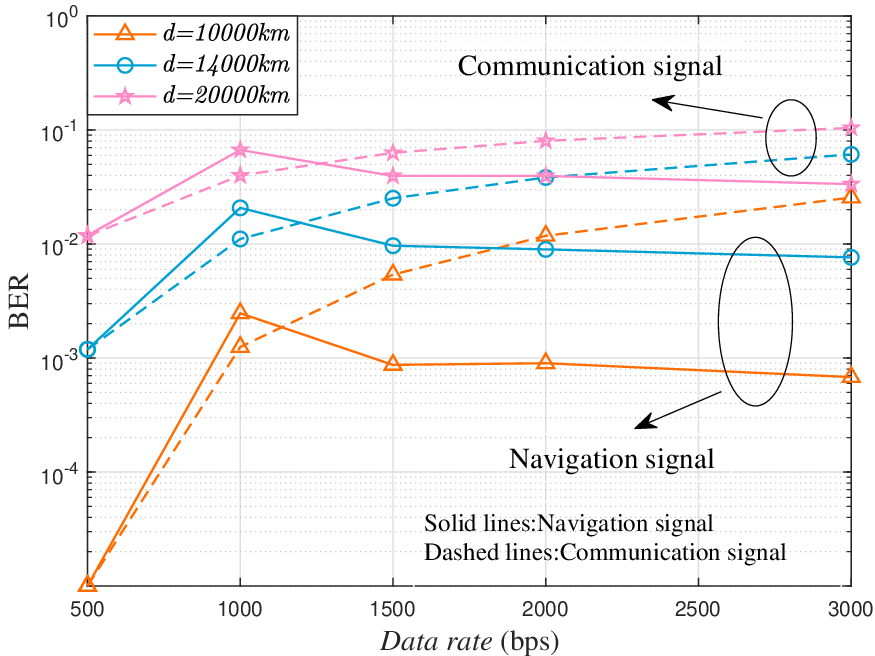}
\caption{The BER of the UO-INAC scenario against the communication data rate.}
\label{CO-Rate_compare}
\end{figure}

\emph{3) Impact of the rate ratio parameter on BER in UO-INAC scenario:} Fig.~\ref{CO-INAC-BER} presents the BER of navigation and communication signals in UO-INAC scenario under varying communication data rates. We can observe that the BER of navigation and communication signals increases as the distance between the satellite and the user increases, which is due to the increased path loss associated with greater distances, leading to a degradation in signal quality. We can also see that as the communication data rate rises, the communication signal requires a higher SNR to maintain an acceptable BER. However, this also results in a more significant susceptibility to noise and interference, which degrades the communication signal's performance. On the other hand, the navigation signal, even at higher data rates, maintains relatively stable performance due to its higher spreading gain, which is less sensitive to the increased data rate. Similar to the MO-INAC scenario, when the rate ratio parameter increases, the BER of navigation signal decreases, while the BER of communication signal increases, which validates our~\textbf{Remark~\ref{remark3}}. Furthermore, the results with estimation errors closely match those under ideal channel estimation, demonstrating the strong robustness of the proposed system.

It can be also seen in Fig.~\ref{CO-Rate_compare} that when the communication signal rate is sufficiently low, the BER performance of communication signal in the UO-INAC scenario is better than that of the navigation signal. The BER performance of communication signal in the UO-INAC scenario is worse than that of the navigation signal when the communication signal rate is high enough. Thus, the results in Figs.~\ref{CO-INAC-BER} and \ref{CO-Rate_compare} demonstrate the importance of managing both communication symbol rate and power allocation to balance the BER performance between the communication and navigation signals.

\begin{figure}[t!]
\centering
\includegraphics[width =3.5in]{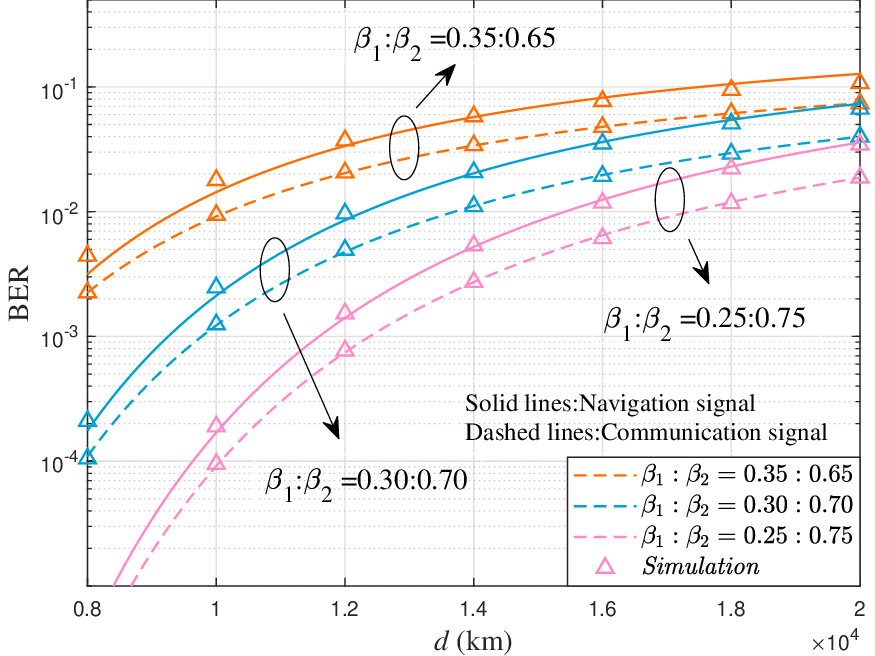}
\caption{The BER of the UO-INAC scenario under different power allocation factors, where the communication data rate is set to 1 kbps.}
\label{CO-beta}
\end{figure}

\emph{4) Impact of the power allocation factor on BER in the UO-INAC scenario:} In Fig.~\ref{CO-beta}, we evaluate the signal BER in the UO-INAC scenario with different power allocation factors. The results indicate that a larger gap between navigation and communication power allocation factors decreases the BER of communication signal, which validates our~\textbf{Remark~\ref{beta CO remark4}}. Notably, as the power allocation factor of the communication signal increases, the BER of both the navigation and communication signals decrease. This suggests that a larger power allocation factor of the communication signal can improve the overall system performance, particularly in terms of the BER of communication signal. Note that the BER performance of the navigation signal improves as the power allocation factor for the communication signal increases. This result may seem counterintuitive at first, but it reflects the interdependence of the signals in the UO-INAC system. Interference in the shared resources is reduced when more power is assigned to the communication signal, benefiting the navigation signal as well, which highlights the importance of optimizing the power allocation to achieve a balance between two signals. However, when the power allocation factor of the navigation signal increases, the BER of the navigation signal also increases, as shown by the pink and blue curves in the figure. This is because the communication signal is assigned less power, which reduces the performance of the communication signal. This performance degradation directly affects the performance of the entire system, resulting in an increase in the BER of the navigation signal, which validates our~\textbf{Remark~\ref{delta CO remark7}}. Thus, there is a trade-off between two signals in terms of power allocation.

\begin{figure}[t!]
\centering
\includegraphics[width =3.5in]{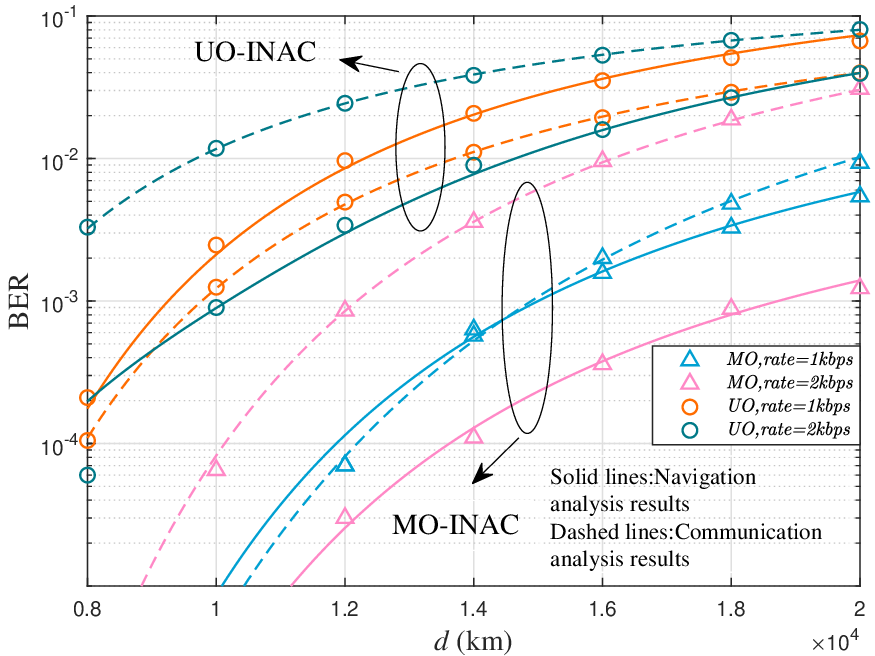}
\caption{Comparison of BER performance for MUC-NOMA-based INAC signals in MO-INAC and UO-INAC scenarios.}
\label{CAO-CO}
\end{figure}

\emph{5) Comparison of BER performance for MUC-NOMA-based INAC signals in MO-INAC and UO-INAC scenarios:} Fig.~\ref{CAO-CO} evaluates the BER performance of the MO-INAC and UO-INAC scenarios. The results show that the BER performance of both the navigation and communication signals in the MO-INAC scenario generally outperforms that in the UO-INAC scenario, suggesting that the MO-INAC system offers a more robust signal in terms of error rates. We can also see that as the distance increases, the gap between the BER performance of two scenarios widens, especially beyond 14,000 km, where the communication signal in the UO-INAC scenario experiences a steep rise in BER. In both the MO-INAC and UO-INAC scenarios, higher rates lead to higher BER, especially noticeable at larger distances. However, the MO-INAC system appears to handle the increased data rate more effectively, maintaining lower BER compared to the UO-INAC system under similar conditions. In conclusion, the MO-INAC configuration is more resilient to path loss and signal degradation with increasing distance, making it the preferred option in MEO satellite communication systems.

\begin{figure}[t!]
\centering
\includegraphics[width =3.5in]{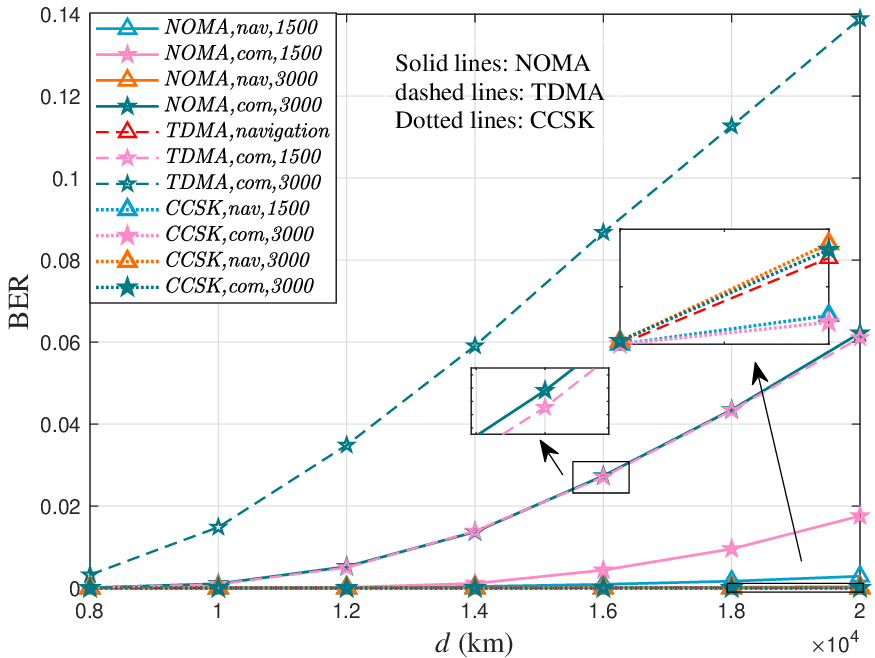}
\caption{Comparison of NOMA with baselines in MO-INAC scenario.}
\label{NOMA-OMA}
\end{figure}

\emph{6) Performance comparison of NOMA with baselines in MO-INAC scenario:} Fig.~\ref{NOMA-OMA} compares the BER performance of the proposed NOMA-based INAC signal with two representative baselines: TDMA and CCSK-based co-modulation. To ensure a strictly fair comparison, all schemes are evaluated under identical conditions, including the same total system bandwidth, the same total observation duration $T$, and the same total transmit energy per symbol. In the TDMA scheme, the navigation and communication signals are transmitted over two orthogonal time slots, each with duration $T/2$. In the CCSK-based scheme, the navigation information is modulated by conventional BPSK, whereas the communication information is embedded into the cyclic shifts of the shared PN sequence.

As observed from Fig.~\ref{NOMA-OMA}, the results indicate a fundamental performance trade-off among the considered schemes. For navigation signal, both the TDMA and CCSK baselines provide slightly better BER performance than the proposed NOMA scheme. This is because TDMA completely eliminates mutual interference through time-domain orthogonality, whereas CCSK fully utilizes the available transmit power and coherent integration interval. By contrast, in the proposed NOMA scheme, the superimposed communication component introduces additional interference during the first-stage navigation decoding, which leads to a slight performance degradation. For communication signal, however, the proposed NOMA scheme outperforms the TDMA baseline. The performance loss of TDMA mainly stems from the orthogonal time allocation, which reduces the effective coherent integration time by half and thus degrades the noise robustness. Meanwhile, the CCSK-based scheme achieves the best communication BER performance among the three schemes, since it avoids both intra-signal interference and error propagation associated with SIC.

However, the proposed NOMA scheme trades the BER penalty for critical system-level advantages that are indispensable for practical INAC systems. In CCSK, the communication data rate is fundamentally limited by the PN sequence length, where transmitting $k$ bits requires $2^k$ distinct cyclic shifts. The proposed NOMA scheme removes the coupling, providing better scalability for high-rate transmission. At the receiver, optimal CCSK detection requires an exhaustive search over $2^k$ possible shifts, leading to exponentially increasing complexity, whereas the proposed NOMA-SIC receiver only incurs linear complexity. In addition, because CCSK embeds data through continuous code-phase shifting, it alters the original ranging structure and reduces the compatibility with legacy navigation receivers. By contrast, the proposed NOMA scheme preserves the original PN phase and maintains ranging integrity. Therefore, despite a slight BER penalty, the proposed NOMA scheme offers a practically attractive trade-off for INAC systems in terms of rate scalability, receiver complexity, and compatibility.

\begin{figure}[t!]
\centering
\includegraphics[width =3.5in]{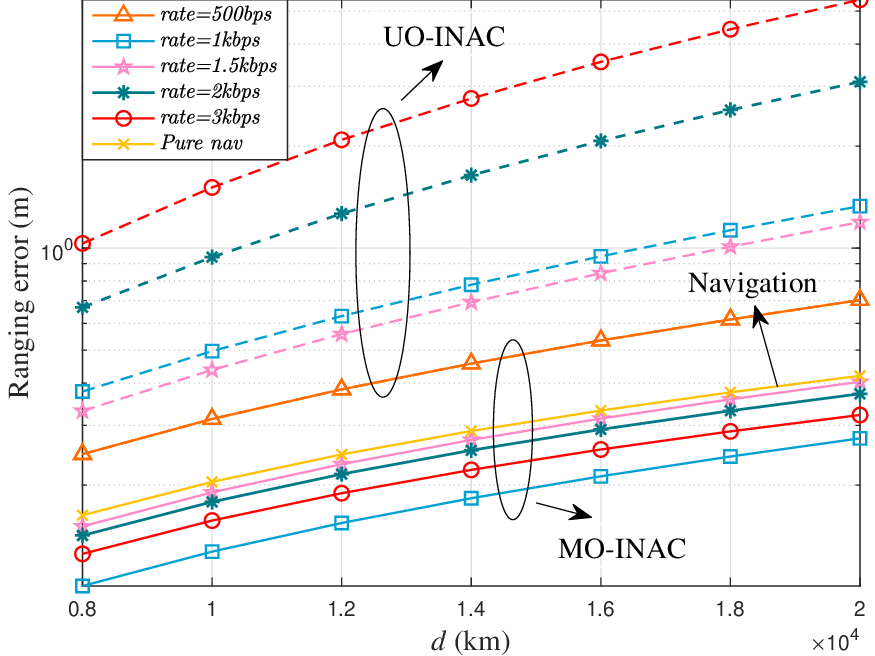}
\caption{The ranging performance in MO-INAC and UO-INAC scenarios.}
\label{ranging}
\end{figure}

\emph{7) Ranging performance analysis:} Fig.~\ref{ranging} depicts the ranging performance of the proposed INAC signal in both MO-INAC and UO-INAC scenarios, compared against a pure navigation baseline. It can be observed that the ranging performance of the MO-INAC scenario is substantially superior to that of the UO-INAC scenario. This is because within the 1 ms correlation integration period, the communication signal contains various bit combinations. In the MO-INAC scenario, the superposition of communication bits with different polarities may cause self-interference cancellation or provide an energy gain to the navigation signal. Therefore, when the communication data rate is greater than 1 kbps, the ranging performance in MO-INAC scenario is better than that of the pure navigation signal. By contrast, in the UO-INAC scenario, the ranging is directly performed by the communication signal. As a result, the ranging accuracy is lower than that of the pure navigation signal, and the ranging performance decreases as the communication data rate increases, which verifies our~\textbf{Remark~\ref{remark-ranging}}.

\section{Conclusions}
We first reviewed the recent advances in INAC signal design. To overcome issues such as the discontinuity of navigation signals in traditional time-division structures and the low spectral efficiency in frequency-division multiplexing structures, we employed the NOMA technique, in which communication and navigation signals are combined. Additionally, at the receiver, we adopted SIC detection techniques to mitigate interference. For characterizing the performance of the MO-INAC and UO-INAC scenarios, the BER of INAC signal and navigation accuracy were derived in closed-form. Future work will focus on combining the proposed signal with other multiple access techniques, such as assigning distinct PN sequences to different user groups and investigating scenarios with multiple concurrent uni-cast streams.

\numberwithin{equation}{section}
\section*{Appendix~A: Proof of Theorem~\ref{Theorem1:navigation signal BER expression}} \label{Appendix:As}
\renewcommand{\theequation}{A.\arabic{equation}}
\setcounter{equation}{0}

The BER of multi-cast signal can be written as:
\begin{equation}\label{BER of cascade signal}
P_{e,mul}^{MO} = \sum\limits_{i = 0}^M {{P_r}\left( {{A_i}} \right){P_r}\left( {{e,mul}|{A_i}} \right)},
\end{equation}
where  ${P_r}\left( {{A_i}} \right)$ represents the probability of ${A_i}$, and ${P_r}\left( {{e,mul}|{A_i}} \right)$ denotes the BER of the navigation signal under the condition that the number of uni-cast symbols with phase 0 are $i$. We assume that the symbols have the same probability, then the probability of $A_i$ can be given by:
\begin{equation}\label{The probability of A_i}
{P_r}\left( {{A_i}} \right) = \frac{{C_M^i}}{2{\sum\limits_{i = 0}^M {C_M^i} }}.
\end{equation}

Based on the PDFs of the despread MO-INAC signal  in \eqref{the PDF of the despread CAO-INAC signal}, ${P_r}\left( {{e_N}|{A_i}} \right)$ can be rewritten as:
\begin{equation}\label{rewritten the BER of the navigation signal under the condition}
\begin{aligned}
{P_r}\left( {{e_N}|{A_i}} \right) &= \frac{1}{2}\int_{ - \infty }^0 {{f_{m,{i_1}}}} (y)dy = \frac{1}{2}\int_0^\infty  {{f_{m,{i_2}}}} (y)dy\\
 &= \frac{1}{2}{\rm{erfc}}\left( {\sqrt {\frac{{{A_i}^2}}{{2{g_{mul}}\sigma _n^2}}} } \right).
\end{aligned}
\end{equation}

By substituting \eqref{The probability of A_i} and \eqref{rewritten the BER of the navigation signal under the condition} into \eqref{BER of cascade signal}, the results in \eqref{navigation signal BER expression in theorem1} can be obtained, and the proof is complete.

\numberwithin{equation}{section}
\section*{Appendix~B: Proof of Corollary~\ref{corollary1:approximated CAO_navigation BER}} \label{Appendix:As}
\renewcommand{\theequation}{B.\arabic{equation}}
\setcounter{equation}{0}

Based on the insights from~\cite{36}, the error function can be further transformed into:
\begin{equation}\label{simple but good approximation for erfc(x)}
{\rm erfc(x)} = \frac{1}{6}{e^{ - {x^2}}} + \frac{1}{2}{e^{ - \frac{4}{3}{x^2}}}.
\end{equation}

Consequently, the summation is strictly dominated by the term with the minimum exponential parameter argument. According to \eqref{A2} and \eqref{The probability of A_i}, the minimum amplitude and the probability of the worst case occurring are given by ${A_0} = {g_{mul}}\sqrt {2{P_s}} \left( {\sqrt {{\beta _1}}  - \sqrt {{\beta _2}} } \right)$ and $C_M^0 / 2^M = 1/2^M$, respectively. Then, the BER can be tightly approximated by the dominant term in the high-SNR regime, which can be written as:
\begin{equation}\label{dominant_term}
\begin{aligned}
&P_{e,mul}^{MO} \approx \frac{1}{{{2^M}}}{\rm{erfc}}\left( {\sqrt {\frac{{A_0^2}}{{2{g_{mul}}\sigma _n^2}}} } \right)\\
&= \frac{1}{{{2^M}}}{\rm{erfc}}\left( {\sqrt {\frac{{{P_s}{g_{mul}}}}{{\sigma _n^2}}} {{\left( {{\beta _1} - {\beta _2}} \right)}^2}} \right).
\end{aligned}
\end{equation}

By substituting \eqref{dominant_term} into the function in \eqref{simple but good approximation for erfc(x)}, we can obtain the closed-form approximation in \eqref{Corro1:approximated CAO_navigation BER}. The proof is complete.

\bibliographystyle{IEEEtran}
\bibliography{IEEEabrv,reference}

\end{document}